\documentclass[review]{elsarticle}

\usepackage{hyperref}

\usepackage{epstopdf}
\usepackage{url}
\usepackage{longtable} 

\journal{Astronomy and Computing}

\begin{document}


\begin{frontmatter}

\title{A hybrid approach to machine learning annotation of large galaxy image databases}


\author{Evan Kuminski 
and Lior Shamir$^{*}$ 
}

\address{Lawrence Technological University \\ 21000 W Ten Mile Rd., Southfield, MI 48075 \\ $^{*}$email: lshamir@mtu.edu}




\begin{abstract}

Modern astronomy relies on massive databases collected by robotic telescopes and digital sky surveys, acquiring data in a much faster pace than what manual analysis can support. Among other data, these sky surveys collect information about millions and sometimes billions of extra-galactic objects. Since the very large number of objects makes manual observation impractical, automatic methods that can analyze and annotate extra-galactic objects are required to fully utilize the discovery power of these databases. Machine learning methods for annotation of celestial objects can be separated broadly into methods that use the photometric information collected by digital sky surveys, and methods that analyze the image of the object. Here we describe a hybrid method that combines photometry and image data to annotate galaxies by their morphology, and a method that uses that information to identify objects that are visually similar to a query object (query-by-example). The results are compared to using just photometric information from SDSS, and to using just the morphological descriptors extracted directly from the images. The comparison shows that for automatic classification the image data provide marginal addition to the information provided by the photometry data. For query-by-example, however, the analysis of the image data provides more information that improves the automatic detection substantially. The source code and binaries of the method can be downloaded through the Astrophysics Source Code Library.


\end{abstract}

\begin{keyword}
Image analysis \sep galaxies \sep galaxy morphology \sep pattern recognition \sep machine learning \sep query-by-example.
\end{keyword}


\end{frontmatter}


\section{Introduction}
\label{introduction}

Modern astronomy has been becoming dependent on data obtained from autonomous digital sky surveys, allowing data-driven and statistical analysis that would not have been possible with manually controlled telescopes \citep{borne2013virtual, djorgovski2013sky,edwards2014astronomy}. As the scales of these databases and the breadth of astronomical pipelines continue to grow, it is expected that efficient and reliable methods for analyzing these data will become crucial for astronomy research.

Perhaps the most basic method of annotating galaxies is through manual analysis performed by scientists \citep{arp1987catalogue,rc3,nair2010catalog,calvi2011padova,efigi}. Clearly, the main downside of that paradigm is its inability to handle the very large databases generated by modern digital sky surveys. To increase the bandwidth of manual annotation, crowdsourcing was used to allow non-expert volunteers to annotate galaxies through a web-based user interface \citep{galaxyzoo1, galaxyzoo2}. However, despite the ability of non-experts to provide useful information about galaxies, the rapidly growing size of these databases makes them far too massive for even a relatively large group of volunteers.

Previously proposed automatic methods for galaxy annotation are based on the application of image processing and computer vision techniques \citep{Abr03,lekshmi2003galaxy,peng2011galfit,simard1998gim2d,baillard2006project,shamir2009automatic,sha11,kum14,dieleman2015rotation,schutter2015galaxy,shamir2013automatic,davis2014sparcfire,hocking2017automatic}, and the application of these algorithms to databases collected by digital sky surveys produced catalogs of morphological information  \citep{hue10,huertas2016mass,shamir2014automatic,kuminski2016computer,ring_catalog}. These catalogs demonstrate that automatic analysis of galaxy images is a practical solution to the problem of annotating large databases of astronomical images.

Another approach to automatic classification and annotation of galaxies is by analyzing the spectroscopic or photometric information produced by the digital sky survey \citep{ball2004,ball2006robust,almeida2010automatic,banerji2010galaxy,vasconcellos2011decision}. That approach does not require direct analysis of the image, but instead the automatic classifier can use a combination of measurements collected by the photometric pipeline of the digital sky survey. The digital sky survey pipelines can provide information that cannot be obtained from the image alone, such as the magnitude of the object in the different bands. However, the information provided by the pipeline is limited to a set of pre-defined measurements that do not contain all information about the galaxy morphology, and full reconstruction of the galaxy morphology using these measurements alone is normally not possible.

Here we combine the analysis of photometric data and computer vision to propose a hybrid method that classifies and detects celestial objects based on both image and photometry data collected by sky surveys. The method combines the pre-defined commonly used photometric measurements with features extracted directly from the images to measure and compare the amount of information that the direct analysis of the images can add to the photometric measurements. Since many modern sky surveys such as Sloan Digital Sky Survey (SDSS) or the Panoramic Survey Telescope and Rapid Response System (PanSTARRS) provide both photometry and image data, such method can be used to perform automatic tasks related to the annotation of celestial objects in large astronomical databases.



\section{Data}
\label{data}

Data were used for two different experiments. The first set of experiments is automatic annotation of galaxy images. That is, given a galaxy image the algorithm annotates the galaxy based on its morphology, and classify visual elements such as the spirality, bulge, number of arms, etc \citep{galaxyzoo2,kum14,dieleman2015rotation}.

The second set of experiments is query-by-example \citep{shamir2016morphology}. Query-by-example allows a researcher to provide the system with a sample of a certain galaxy of interest, and the system returns a list of objects that are the most similar to the query object, allowing the studying of that object using more than one galaxy of the same type. Clearly, the availability of a group of similar objects allows statistical analysis of their characteristics, such that $N>1$.

\subsection{Data for the automatic annotation of galaxies}

The data used in the experiment are taken from SDSS, and annotated by Galaxy Zoo 2 \citep{galaxyzoo2}, as thoroughly explained in \citep{kum14}. The images are JPEG images of dimensionality of 120$\times$120 pixels, downloaded through SDSS's Catalog Archive Server (CAS). The JPEG images are generated in the CAS database by combining the FITS images taken in the i, r, and g bands, invoked in each request sent to the ImgCutout service of CAS \citep{lupton2004preparing}. The images are generated after applying de-noising and several filters that can change the images, and ignores the b and z bands, and therefore the JPEG images might not contain all information contained in the raw FITS images. However, these JPEG images normally contain complete information that allows to visually determine the morphology of the galaxy, and therefore these images are the most informative images to be used by machine learning systems compared to using the individual bands.

The galaxies in the Galaxy Zoo 2 catalog have different angular sizes, so the galaxy images are downloaded in an iterative process starting with a scale of 0.1'' per pixel. The image is separated to foreground and background pixels using the Otsu binary transform \citep{otsu1979threshold}, and the scale is increased by 0.05'' per pixel until no more than 40 foreground pixels are located on the edge of the frame. The iterative scaling of the images ensures that the object fits inside the frame.

The data used in this study are the original 245,609 Galaxy Zoo 2 images, and the morphological annotations associated with each image. Since each galaxy is annotated by 44  non-expert participants on average, the resulting annotation of each galaxy and each question is determined statistically by the distribution of the votes. Naturally, when the level of agreement between the voter gets higher, the annotation can be considered more likely to be correct \citep{galaxyzoo1}.

The distribution of the votes for each question provided the ground truth data for the experiments. 
To reduce possible noise, only questions that their answers in the context of a certain galaxy reached a certain threshold of agreement were used. When the distribution of the answers for a certain galaxy does not reach that threshold, the galaxy is rejected and not used in the experiment. A detailed description of the data can be found in \citep{kum14}. 

The size of the dataset for each question changes with the agreement threshold. The sizes of the classes and the total size of the dataset of each question are specified in \citep{kum14}, and also in Table~\ref{dataset_sizes}. To avoid bias due to differences in the sizes of the classes in the GZ2 sample, the size of each class equals to the number of galaxies in the smallest class \citep{kum14}.

\begin{table*}[h]
\scriptsize
\begin{tabular}{|l|c|c|c|c|c|c|c|}
\hline
Question & $>$50\% & $>$60\% & $>$70\%	& $>$80\% & $>$90\%	& $>$95\% & $>$97\%   \\
\hline
1          & 25000 & 25000 & 25000 & 25000  & 19693  & 6635  & 2332 \\
            &    (241679) & (213890) & (181715) & (132748)  & (50946) & (19248) & (12193)  \\
2 & 10367 & 6705 & 3955 & 2003 & 645 & 225 & 127  \\
   & (61515) & (53058) & (41193) &  (27891) & (18325) & (10154) & (6761) \\
3 & 13993 &	9964 & 6506 & 3720 & 1399 & 386 & 171  \\
   & (55488) & (46176) &  (37181) &  (27788) & (17646) & (9913) & (6629) \\
4 & 9846 & 4334 & 1522 &	323 &	11 & 1 & 0 \\ 
   & (55396)  & (42130) & (32119) & (23681) &  (15106) & (8421) & (5479) \\
5 & 13028 &	5866 & 510 & 110 & 3 & 0 & 0 \\
  & (42780) &  (24385) & (10583) & (2499) & (143) & (9) & (2) \\
6 & 22889 & 15791 & 9921 & 5369 & 1957 & 691 & 417 \\
   & (242291)	& (224645) & (202509) & (170574) & (115692) &  (65638) &  (46159) \\
7 & 24203 & 	16442 & 8593 & 2117 & 103 & 6 & 4 \\
  & (172761) &  (138128) & (100328) &  (55774) & (9545) & (961) & (221) \\
8 & 37 & 9 & 1  & 0 & 0 & 0 & 0 \\ 
   &  (15219) & (7483) & (3078) & (1008) & (163) & (18) & (11) \\ 
9 & 97 & 44  & 18 & 8 & 3 & 0 & 0 \\
   & (9272) & (4860) & (2076) & (562) & (48) & (4) & (1) \\
10 & 5471 &	3371 & 1337 & 119 & 6  & 0 & 0 \\
   & (33536) &  (16289) & (6332) & (2061) & (469) & (118) & (51) \\
11 & 226 & 120 & 58  & 24 & 0 & 0 & 0 \\
   &  (21814) & (14966) & (10900) & (8490) & (5948) & (3568) & (2343)  \\
\hline
\end{tabular}
\caption{The size of each class used for automatic annotation and the total number of galaxies (in parentheses) for each GZ2 agreement threshold \citep{kum14}.}
\label{dataset_sizes}
\end{table*}

\subsection{Data for galaxy query-by-example experiments}

The set of galaxies used in the query-by-example experiments is described in \citep{shamir2016morphology}. That data is also taken from SDSS, downloaded in the same manner, and provided several datasets. The first dataset contains galaxies classified into spiral or elliptical galaxies \citep{kuminski2016computer}. In a universe that contains just early-type galaxies, a spiral galaxy would be considered ``peculiar'', and therefore a small set of spiral galaxies can be combined with a larger number of elliptical galaxies, and then each of the spiral galaxy images can be used as a query galaxy. The performance can be measured by the number of spiral galaxies among the total number of galaxies returned by the queries.

The first dataset contains 100 spiral galaxies and 100 elliptical galaxies taken from \citep{kuminski2016computer}. The galaxies were visually inspected, and used also in \citep{shamir2016morphology}. The elliptical and spiral galaxies were also combined with 20 ring galaxies and 20 galaxy pairs  \citep{shamir2016morphology} as the query galaxies.

In addition to the smaller datasets, another dataset of 4,000 images of galaxies classified as spiral and 4,000 images of galaxies classified as elliptical were used in combination with ring and galaxy pairs. These elliptical and spiral galaxies are also taken from the catalog of SDSS galaxies classified by their broad morphology \citep{kuminski2016computer}.



\section{Methods}
\label{methods}

\subsection{Morphological features}
\label{morph}

The morphological features used in this experiment are based on the morphological features of the Wndchrm scheme \citep{shamir2008wndchrm,shamir2013wnd}, which demonstrated its ability to analyze galaxy images \citep{shamir2009automatic,shamir2013automatic,kum14}. The method works by first extracting a large set of numerical image content descriptors, including texture features such as the Tamura, Gabor, and Haralick textures, pixel intensity distribution such as multi-scale histograms and first four moments, shape features such as edge and object statistics, polynomial decomposition such as Chebyshev statistics and Zernike polynomial, Radon features, fractals, and the Gini coefficient \citep{Abr03}. One of the unique elements of Wndchrm is that the numerical image content descriptors are computed not just from the original image, but also from transforms of the image. The transforms used by the algorithm are the Fourier transform, Chebyshev transform, Wavelet transform (Symlet 5), and edge magnitude transform,in addition to combinations of these transforms. These morphological features were used for both the automatic annotation and the query-by-example experiments. A detailed description of the  numerical image content descriptors can be found in \citep{shamir2008wndchrm,shamir2009automatic,shamir2013automatic,shamir2009knee,shamir2014classification}, and the source code is publicly available \citep{shamir2013wnd}.

\subsection{Photometric features}
\label{phot}

Photometric data were obtained from SDSS Data Release 7 \citep{abazajian2009seventh}, and included the 453 columns of the PhotoObjAll table. To avoid incomplete data, objects that had missing or bad measurement values of -1000 or -9999 were rejected from the experiment, reducing the Galaxy Zoo 2 dataset to 138,232 objects with complete photometric values in the PhotoObjAll table. These features are added to the morphological features to create a single feature vector that contains both morphological and photometric features. For instance, the 453 photometric features are added to the 2,883 morphological features to create a feature vector of the size of 3,336 features.

While not all fields in the PhotoObjAll table are necessarily informative (e.g., the object ID), the use of feature selection as will be described in Section~\ref{classification} automatically removes features that do not provide meaningful information.

\subsection{Pattern recognition}

\subsubsection{Classification}
\label{classification}

The classification is based on the Weighted Nearest Distance (WND) scheme \citep{orlov2008wnd,shamir2008wndchrm}. First, all morphological and photometric features are normalized to the interval [0,1], and assigned with Fisher discriminants scores \citep{bishop2006pattern}. The 85\% of the features with the lowest Fisher discriminant scores are rejected, and the remaining features are used by the Weighted Nearest Distance (WND) classifier such that the Fisher discriminant scores are used as weights \citep{orlov2008wnd,shamir2008wndchrm}. 

The WND classifier is based on the distance shown in Equation~\ref{distance}

\begin{equation}
d(x,c)=
{\frac 
{\sum_{t \in T_c}[\sum_{f=1}^{|x|}W_f^2(x_f-t_f)^2]^{p}}  
{| T_c |}}
\label{distance}
\end{equation}
where {\it T} is the entire training set, $T_c$ is the training set of class $c$, $t$ is a feature vector from $T_c$, {\it x} is the feature vector of the galaxy image being classified, $|x|$ is the size of feature vector, $x_f$ is the value of image feature {\it f}, $W_f$ is the Fisher discriminant score of feature {\it f}, $|T_c|$ is the number of training samples of class $c$, $d(x,c)$ is the computed distance from a given sample {\it x} to class {\it c}, and $p$ is the exponent, which is set to -5 as described in detail with empirical results in \citep{orlov2008wnd}. Naturally, the class $c$ that has the shortest distance to the sample $x$ is determined as the predicted class.

The performance is measured simply by the number of correct annotations made by the algorithm, divided by the total number of galaxies being annotated. An annotation is considered correct if the annotation made by the computer is the same as the annotation made by the majority of the Galaxy Zoo 2 citizen scientists. The ``majority'' threshold is dynamic, and different thresholds are tested for each question as will be described in Section~\ref{results}.

\subsubsection{Query by example}
\label{qbe}

The pattern recognition for the query-by-example method is described in \citep{shamir2016morphology}. The method is based on the same numerical image content descriptors mentioned in Sections~\ref{morph} and~\ref{phot}, but measures the distances between the query object and each of the objects in the dataset using weighted Euclidean distance or by Earth Mover's Distance \citep{rubner2000earth}, after weighting the descriptors by their entropy \citep{shamir2016morphology}. The source code of the method is publicly available \citep{shamir2017udat}. As shown in \citep{shamir2016morphology}, the best performance of the query-by-example algorithm is achieved when using entropy weights and Earth's Movers Distances. 

The performance of the query by example is measured by using two sets of galaxies -- a database set and a query set. The database set contains galaxies that are the typical non-peculiar galaxies in the database, and the query set contains galaxies of interest that the algorithm attempts to find among the galaxies in the database set. 

The performance evaluation is done by merging $P$ galaxies from the query set with $Q$ galaxies from the database set. Then, one of the galaxies from the query set is selected as the query galaxy, while a subset of the remaining $P$-1 galaxies is merged with the $Q$ database galaxies, and the algorithm returns the $R$ galaxies that the algorithm determines are the most similar to the query galaxy image. The process is repeated $P$ times such that all galaxies in the query set are used once as the query galaxy. 

The performance is determined by Equation~\ref{performance}

\begin{equation}
\label{performance}
\frac{\Sigma_{p=1}^{|P|} \Sigma_{r=1}^{|R_p|} (R_{p_r} \in P \wedge R_{p_r} \neq p)} {|P|} ,
\end{equation}
where $R_p$ is the set of galaxies returned by the algorithm as the most similar to the query galaxy image $p$. 

The rank is defined as the size of the set $R_p$. If a galaxy from the query set $P$ is present among the $R$ galaxies the query is considered a hit. The performance is determined by the hit rate, which is the average number of galaxies of the query set $P$ among the $R$ galaxies returned by the method in each of the $|P|$ queries it attempts. That process is repeated iteratively such that in each run a different galaxy $p$ is used as the query galaxy, and the performance is evaluated by the average number of galaxies of the query set among the top R galaxies returned by the query. 

The method is not expected to be fully accurate, and therefore it is expected that the set $R$ returned by the method can also contain many galaxies that are not of the same type as the query galaxy. However, the purpose of the method is to reduce the data and provide a smaller dataset in which the frequency of galaxies similar to the query galaxy is much higher compared to their population in the database \citep{shamir2016morphology}.

\section{Results}
\label{results}

\subsection{Classification}
\label{classification_results}

To quantify the efficacy of combining morphological data and photometric data, we ran a series of tests using the results of Galaxy Zoo 2 data described in Section~\ref{data} to obtain training and test data. By changing the threshold of agreement we controlled the trade-off between the size of the training set and the cleanness of the data. Setting a high threshold for the vote fraction leads to a smaller but cleaner dataset, as higher agreement rate between the voters leads to cleaner annotations. Lower threshold provides a larger dataset in which more objects are more likely to have been misclassified. This process is explained in greater detail in \citep{kum14}.

For each of the training sets that were generated, the algorithm was tested with the three feature sets described in Section~\ref{methods}: the photometric features, the morphological features, and a combination of both morphological and photometric feature sets. This was done to verify that the algorithm could draw useful information from the photometric data, and to then evaluate how the combined feature set is compared to the morphological and photometric feature sets.


As done in \citep{kum14} and mentioned in Section~\ref{methods}, each question in GZ2 is tested by separating the data based on the degree of agreement among the citizen scientists who annotated that question. That was repeated for each degree of agreement, so that only galaxies that were annotated with that agreement level or higher are used, and the rest of the galaxies are rejected from the experiment.

\begin{figure*}[p]
\includegraphics[scale=0.58] {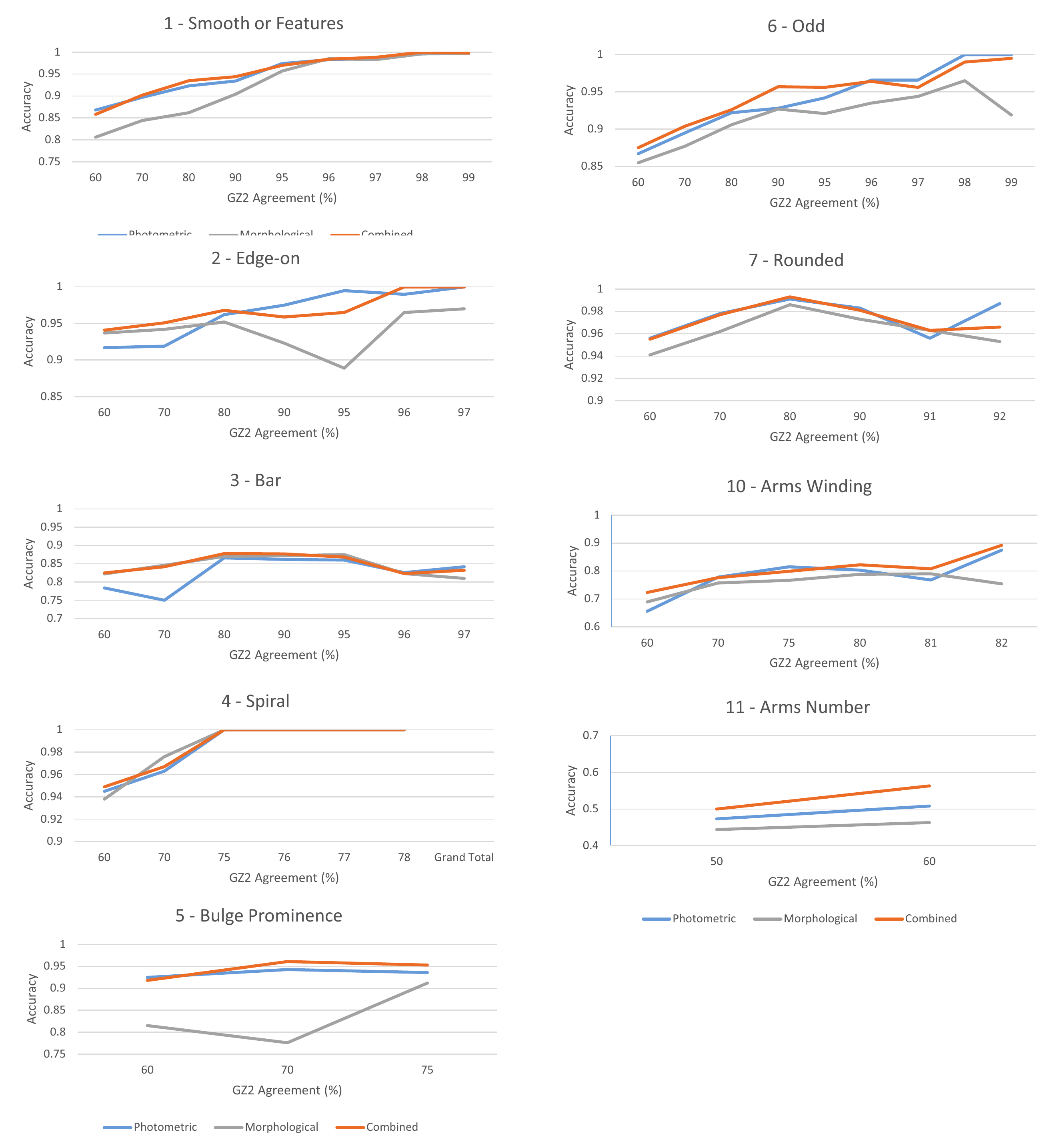}
\caption{Comparison of the classification accuracy for different feature sets when applied to different question in Galaxy Zoo 2.}
\label{euc_distance}
\end{figure*}

Figure~\ref{euc_distance} shows the automatic classification accuracy when the galaxies are separated into classes based on the manual annotations of the different questions in Galaxy Zoo 2. That is, the manual annotations of the galaxies obtained through the Galaxy Zoo 2 campaign are used as ground truth for training and testing the machine learning algorithm. When the Galaxy Zoo 2 agreement threshold of the annotation is increased, less galaxies are used for training. That can lead to a decrease in classification accuracy, as the performance of machine learning algorithms depends on the size of the training set. However, including galaxies on which the agreement between the citizen scientists is higher leads to a cleaner and more consistent dataset \citep{galaxyzoo1}, which can compensate for the smaller size of the training data.

As the graphs show, the classification accuracy for most questions using the photometric features was similar to the classification accuracy achieved using the morphological features, but in seven of the nine cases using the photometric features provided higher accuracy in comparison to the morphological features. When combining the photometric and morphological features, in most cases the classification accuracy is improved for the different questions and different threshold values compared to using either the photometric features or the morphological features alone. Perhaps the only notable exception is question 2 (``Could this be a disk viewed edge-on?''), in which the photometric features outperform the combined feature set when using galaxies that were annotated by the Galaxy Zoo 2 volunteers in agreement level of 90--95\%. It should be noted that while the answer to that question is boolean (``yes'' or ``no''), there are many in-between cases based on the position of the galaxy in comparison to Earth. Question 8 (odd features) and 9 (bulge shape) of Galaxy Zoo 2 were not used due to the low number of sample classifications.

The algorithm selects the most informative features automatically based on their Fisher discriminant scores. The top 30 descriptors and their Fisher discriminant scores when using the morphological features, the photometric features, and the combined features used in Question 1 (whether the galaxy is round and smooth) are specified in Table~\ref{features_rank} ranked by their Fisher discriminant scores. The most informative photometric feature in the context of Question 1 is rho (log size for surface brightness in the i band). Among the morphological features, most of the features were texture features (e.g, Haralick) and Fractals.

\begin{table*}[h]
\tiny
\begin{tabular}{|l||l|l|l|}
\hline
Rank & Morphological features & Photometric features & Combined feature set \\
\hline
1 & Haarlick Texture 5 (Fourier Chebyshev) : 3.52 & rho: 4.34  & rho: 4.35 \\
2 & Fractal bin 3 (Chebyshev) : 3.194908 &  deVMag\_u: 3.77  & deVMag\_u: 3.77 \\ 
3 & Haarlick Texture 10 (Fourier Wavelet) : 3.15 & petroR50\_u: 3.42 & Haarlick Texture 5 (Fourier Chebyshev): 3.52 \\
4 & Haarlick Texture 0 (Fourier Wavelet) : 3.09  & deVMag\_g: 3.4 & deVMag\_g: 3.41  \\
5 & Haarlick Texture 18 (Fourier Wavelet) : 2.95 &  expRad\_g: 3.31 & petroR50\_u: 3.38 \\
6 & Fractal bin 11 (Chebyshev Wavelet) : 2.91  & petroR50\_g: 3.31 & expRad\_g: 3.32 \\
7 & Haarlick Texture 20 (Fourier Wavelet) : 2.84 &  expMag\_u: 3.3  & expMag\_u: 3.31 \\
8 & Fractal bin 6 (Chebyshev Wavelet) : 2.84  & deVRad\_u: 3.12 & petroR50\_g: 3.27  \\
9 & Haarlick Texture 8 (Fourier Wavelet) : 2.8  & petroR50\_r: 3.12 & Fractal bin 3 (Chebyshev): 3.22 \\
10 & Fractal bin 10 (Chebyshev Wavelet) : 2.78 &  petroR50\_i: 3.05 & Haarlick Texture 10 (Fourier Wavelet): 3.19 \\
11 & Haarlick Texture 4 (Fourier Chebyshev) : 2.75 &  expRad\_r: 3.02 & Haarlick Texture 0 (Fourier Wavelet): 3.16 \\
12 & Haarlick Texture 14 (Fourier Wavelet) : 2.73  & petroR50\_z: 2.99  & deVRad\_u: 3.15  \\
13 & mean (Fourier Wavelet) : 2.726157  & deVRad\_g: 2.98 & petroR50\_r: 3.1  \\
14 & Fractal bin 5 (Chebyshev Wavelet) : 2.72  & petroRad\_g: 2.91 & petroR50\_i: 3.02 \\
15 & Haarlick Texture 12 (Fourier Wavelet) : 2.72  & petroMag\_u: 2.89 & deVRad\_g: 3 \\
16 & Fractal bin 15 (Chebyshev Wavelet) : 2.7  & expRad\_i: 2.89 & expRad\_r: 2.99  \\
17 & Fractal bin 4 (Chebyshev Wavelet) : 2.68  & petroRad\_r: 2.84  & Haarlick Texture 18 (Fourier Wavelet): 2.99  \\
18 & MultiScale Histogram bin 1 (Wavelet) : 2.67  & expRad\_u: 2.83  & petroR50\_z: 2.97  \\
19 & Fractal bin 16 (Chebyshev Wavelet) : 2.67  & dered\_u: 2.81 & Fractal bin 11 (Chebyshev Wavelet): 2.89  \\
20 & Zernike bin 2 () : 2.658163  & expMag\_g: 2.79 & petroRad\_g: 2.89 \\
21 & gini coefficient (Fourier Chebyshev) : 2.63  &  petroRad\_i: 2.79 & Haarlick Texture 20 (Fourier Wavelet): 2.89 \\ 
22 & Zernike bin 12 (Wavelet) : 2.630277  & modelMag\_u: 2.74 & expRad\_i: 2.89  \\
23 & Fractal bin 12 (Chebyshev Wavelet) : 2.63  & modelMag\_u: 2.74 & petroMag\_u: 2.89 \\
24 & Fractal bin 19 (Chebyshev Wavelet) : 2.62  & deVMag\_r: 2.56 & expRad\_u: 2.84 \\
25 & gini coefficient (Chebyshev Fourier) : 2.62  & petroRad\_u: 2.47 & dered\_u: 2.82  \\
26 & Fractal bin 14 (Chebyshev Wavelet) : 2.61  & petroMag\_g: 2.41 & Fractal bin 6 (Chebyshev Wavelet): 2.82  \\
27 & Haarlick Texture 24 (Fourier Wavelet) : 2.61  & deVRad\_r: 2.41 & Haarlick Texture 8 (Fourier Wavelet): 2.82 \\
28 & CombFirstFourMoments 11 (Wavelet) : 2.61  & isoA\_u: 2.4 & petroRad\_r: 2.81  \\
29 & Zernike bin 0 (Edge Transform) : 2.6  & expRad\_z: 2.40 & expMag\_g: 2.8  \\
30 & Fractal bin 7 (Chebyshev) : 2.58  &  petroRad\_z: 2.38 & petroRad\_i: 2.79 \\
\hline
\end{tabular}
\caption{The most informative morphological and photometric features and their Fisher discriminant scores. These features are used for question 1, which is whether the galaxy is round and smooth.}
\label{features_rank}
\end{table*}

\begin{figure*}[h]
\includegraphics[scale=1.00]{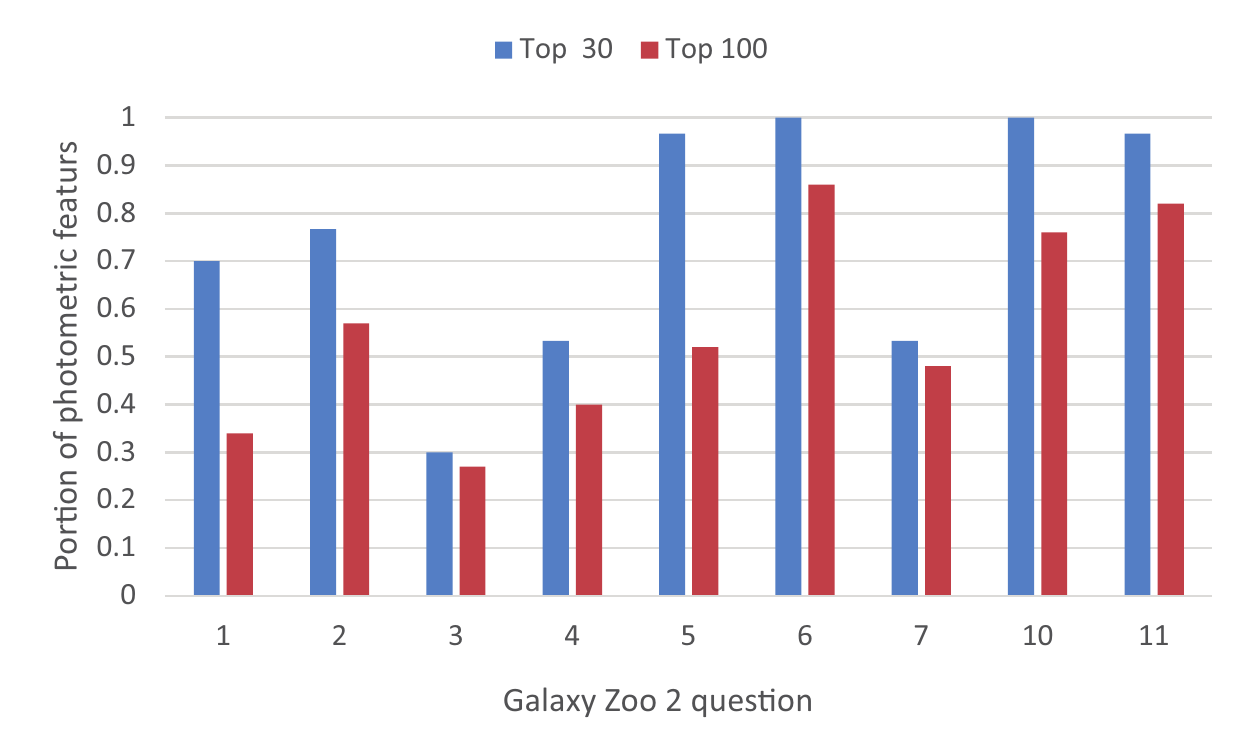}
\caption{Portion of photometric features among the 30 and 100 most informative features for the analysis of the different Galaxy Zoo 2 questions.}
\label{amount_features}
\end{figure*}

The table shows that among the top 30 features in the combined feature set, 21 were photometric features. Among the 100 most informative features 34 were photometric features, while 66 were the morphological features. Figure~\ref{amount_features} shows the amount of photometric and morphological features among the 30 and 100 most informative descriptors for each of the Galaxy Zoo 2 questions that provided the ground truth information for the experiments. The figure shows that the most informative features can be mostly photometric or morphological based on the specific question.

As Table~\ref{features_rank} shows, the photometric features that are assessed as the most informative and have the highest impact on the analysis are not features that reflect the morphology of the galaxy directly, but are related to size (e.g. Petrosian radius) and brightness (e.g., model magnitude). These features correlate with the different types of galaxies, and their combination provides patterns that allow the identification of the morphology of the galaxy without measuring it directly. On the other hand, the machine vision-based morphological features are features such as fractals and textures, that are clearly driven directly by the shape of the galaxy and reflect its the morphology.

\begin{figure}[h]
\includegraphics[scale=0.60]{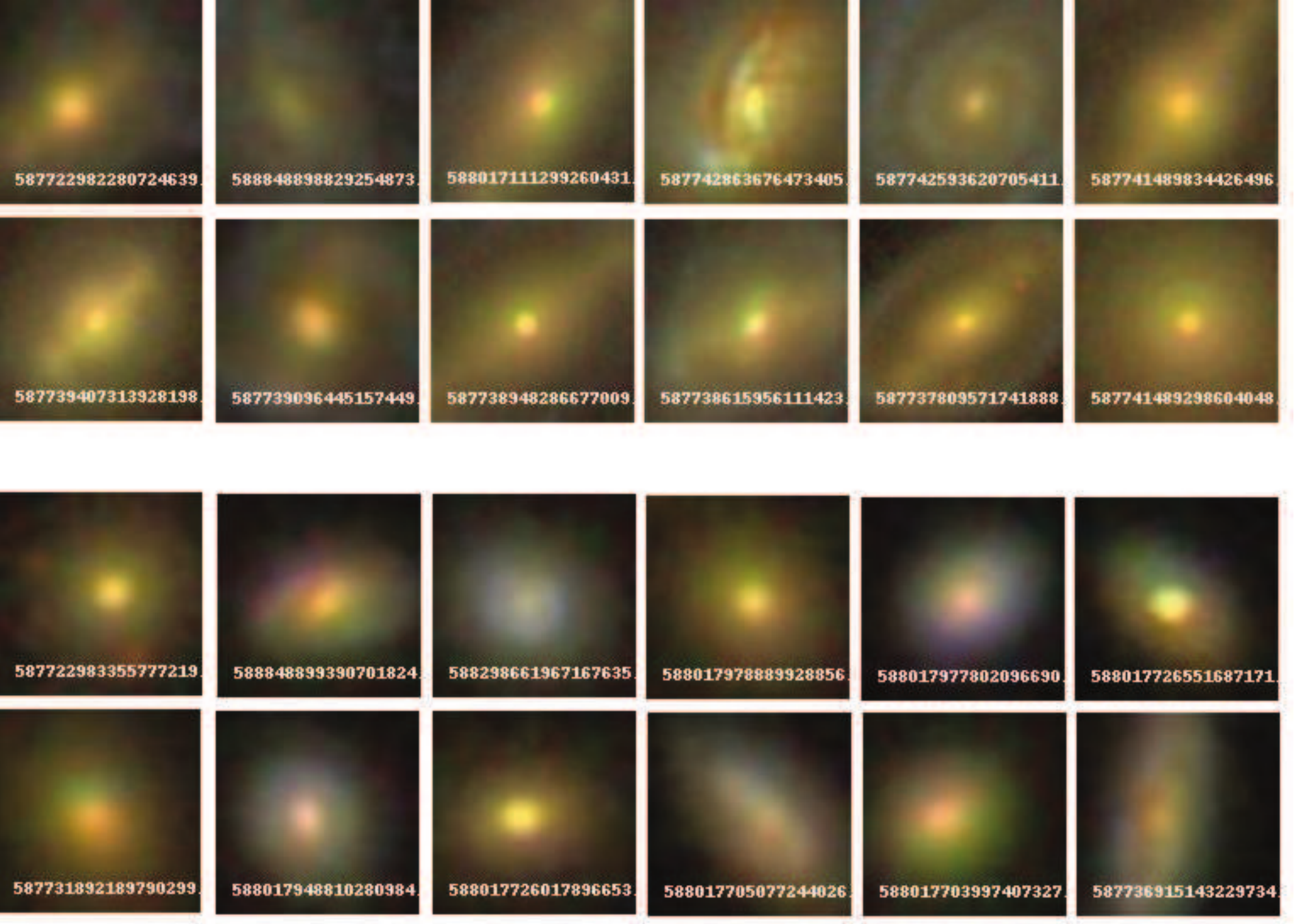}
\caption{Examples of false detections of spiral and elliptical galaxies. The top two lines are galaxies that were classified by Galaxy Zoo 2 as ``superclean'' spiral but were classified by the method as elliptical, and the bottom line shows galaxies that were classified by Galaxy Zoo 2 as elliptical but the method identified them as spiral. The DR8 IDs were added to the images to allow the identification of the specific galaxies, but were not part of the original galaxy images.}
\label{false_detection}
\end{figure}

Figure~\ref{false_detection} shows examples of false positives and false negative classifications, such that the ground truth is the Galaxy Zoo 2 ``superclean'' classifications. These false detection show that galaxies identified as ellipticals by the method can in fact be spiral galaxies. On the other hand, galaxies identified as elliptical by human annotators might sometimes have spiral features in them, as clearly seen in galaxy 587736915143229734. These differences between human and machine classification are aligned with the observation that human classifiers tend to misidentify spiral galaxies and annotate them as elliptical \citep{dojcsak2014}.

\subsection{Query by example}
\label{qbe_results}



\begin{figure}[h]
\includegraphics[scale=0.90] {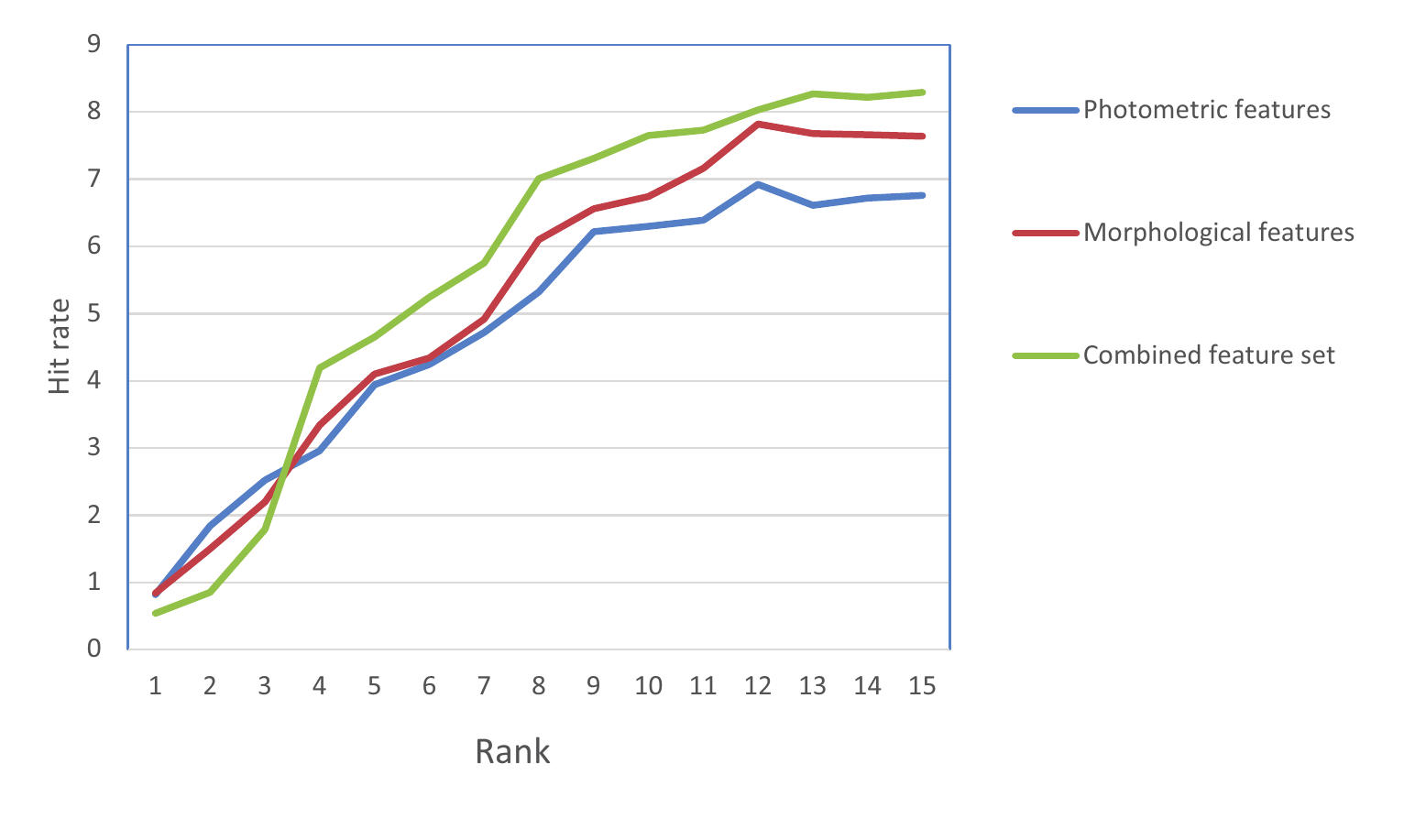}
\caption{Hit rate of the query by example algorithm when using 100 spiral galaxies as the database class and 10 elliptical galaxies as the query galaxies. The experiment was repeated with the morphological features alone, the photometric features alone, and both photometric and morphological features.}
\label{elliptical_in_spiral_100}
\end{figure}

\begin{figure}[h]
\includegraphics[scale=0.90] {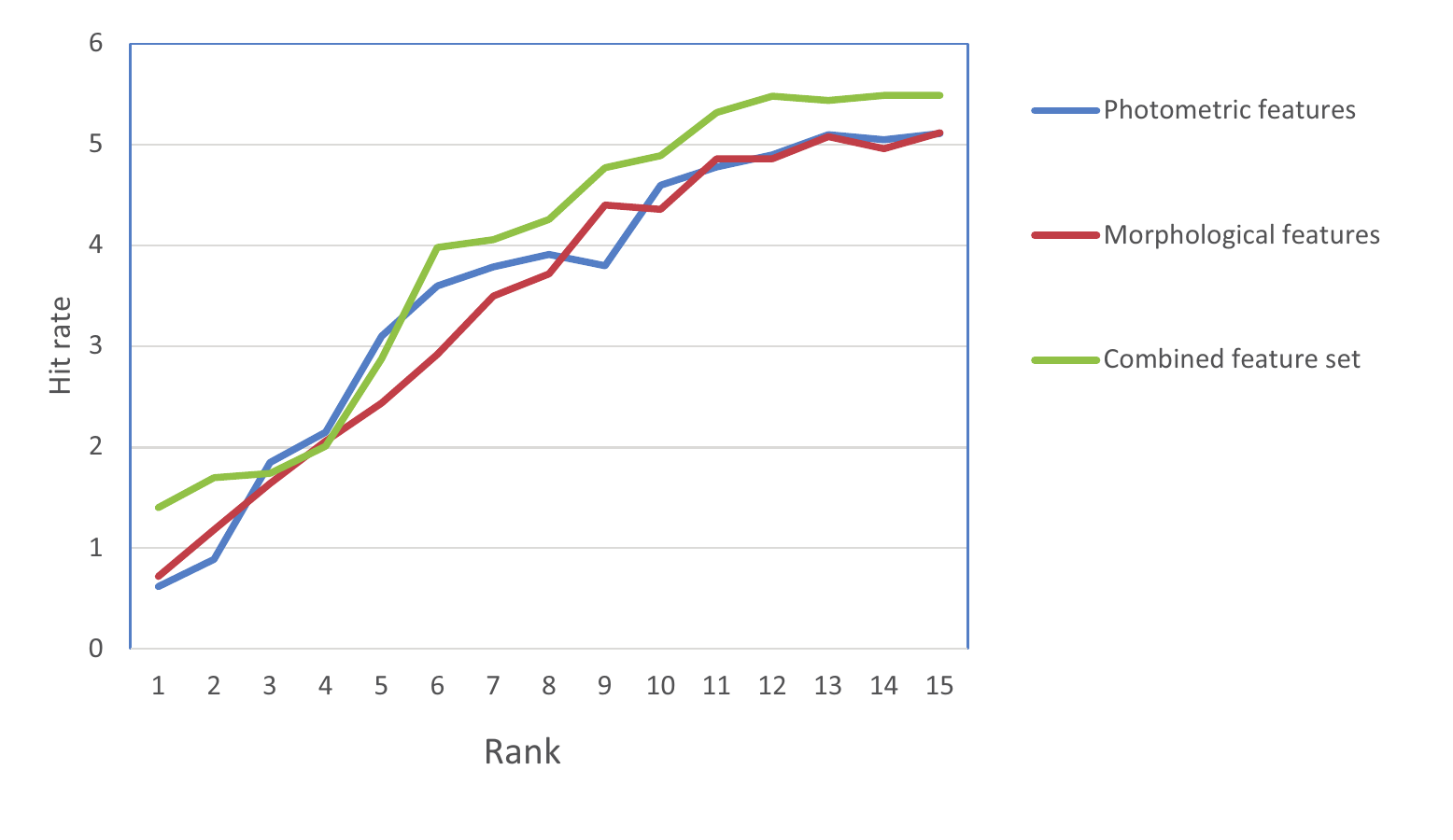}
\caption{Hit rate of the query by example algorithm with the different feature sets when using 100 elliptical galaxies as the database class and 10 spiral galaxies as the query galaxies.}
\label{spiral_in_elliptical_100}
\end{figure}

\begin{figure}[h]
\includegraphics[scale=0.90] {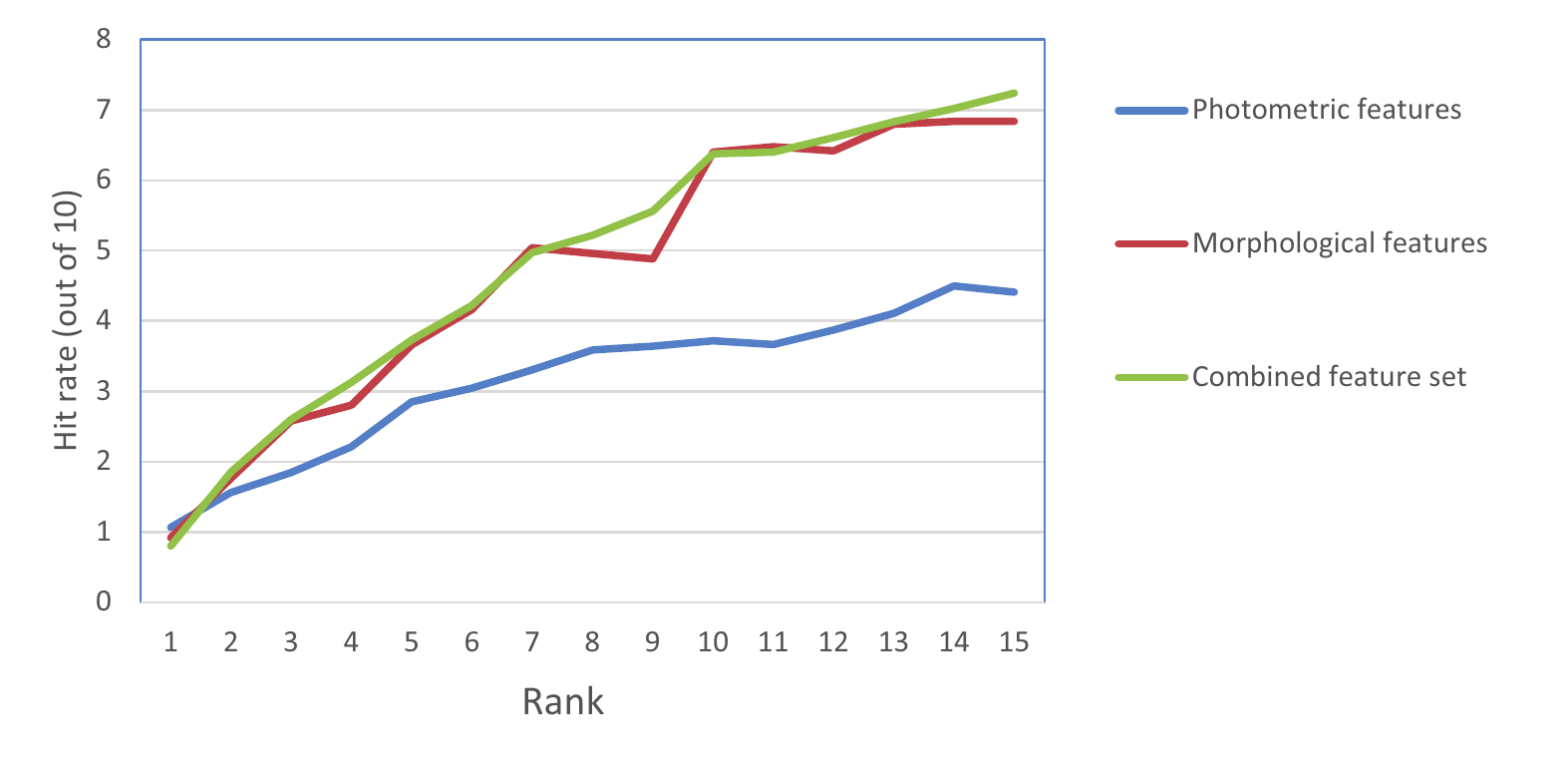}
\caption{Hit rate of the query-by-example algorithm when 10 ring galaxies are combined with 100 elliptical galaxies. The algorithm is performed with the morphological features, the photometric features, and the combined feature set.}
\label{rings_in_ellipticals}
\end{figure}

\begin{figure}[h]
\includegraphics[scale=0.90] {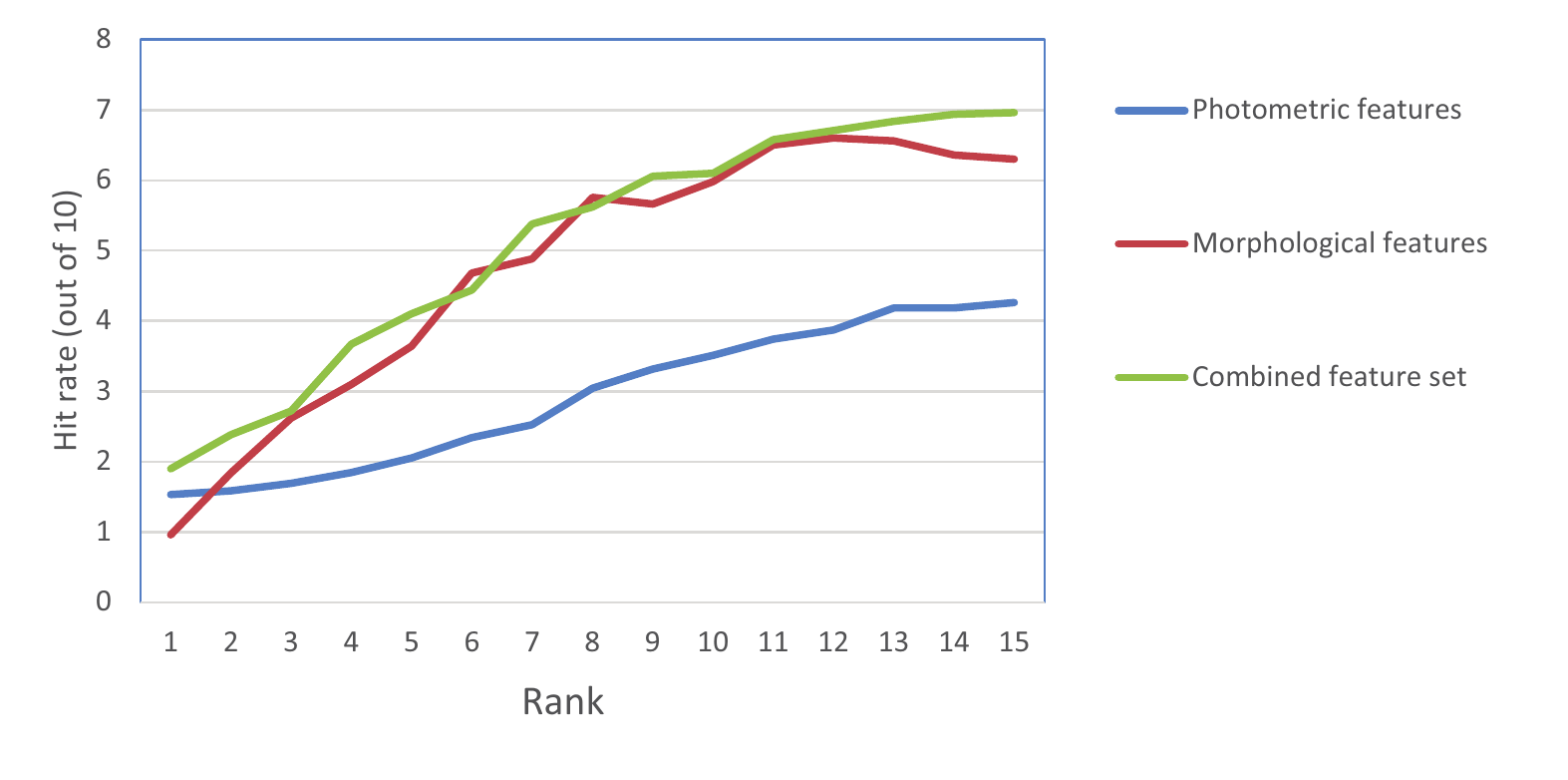}
\caption{Hit rate of the query by example algorithm of the different feature set when using 100 spiral galaxies and 10 ring galaxies, and the query galaxy is a ring galaxy.}
\label{rings_in_spiral}
\end{figure}

\begin{figure}[h]
\includegraphics[scale=0.65] {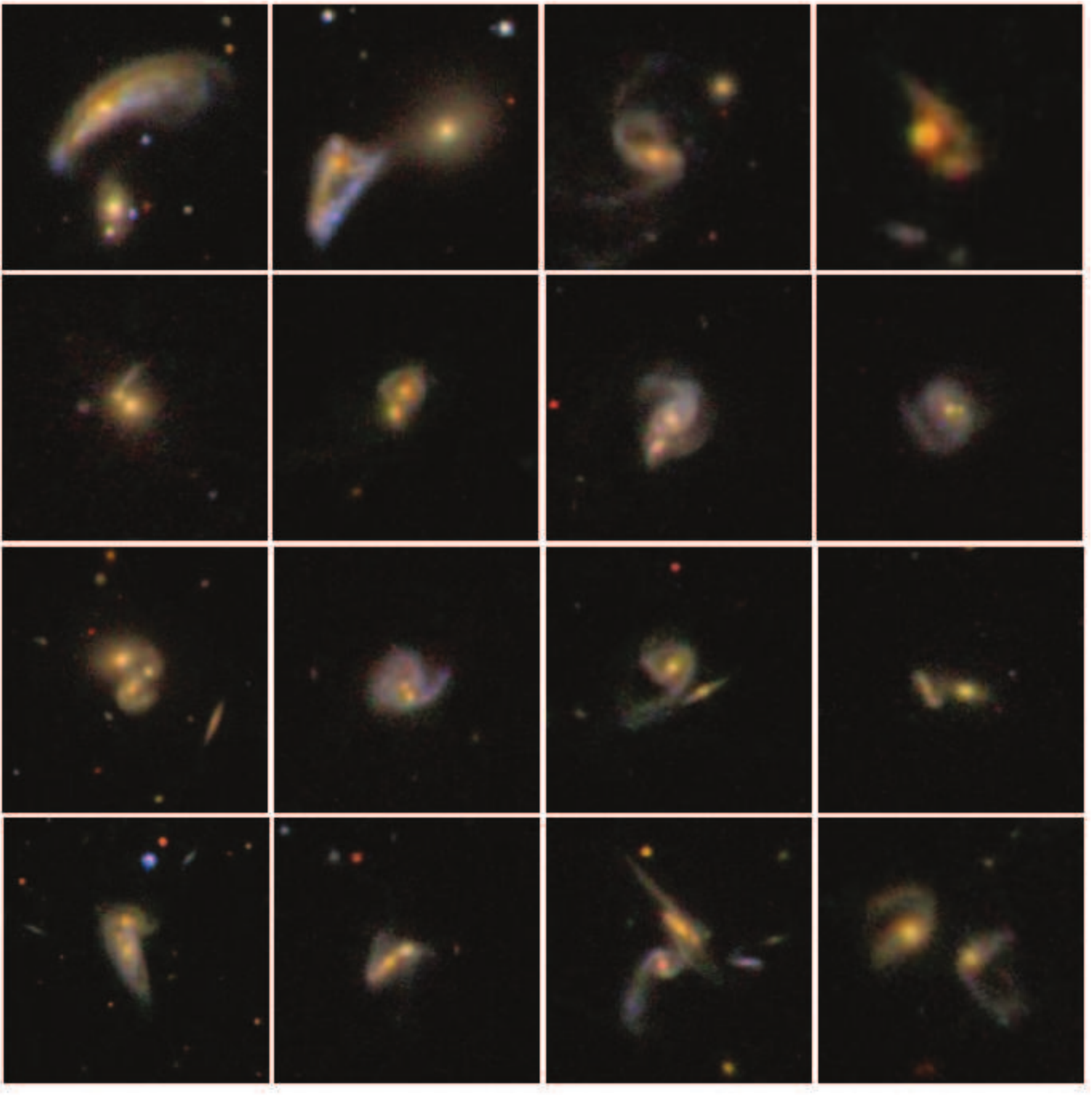}
\caption{Tidally distorted galaxy pairs taken from the catalog of automatically detected peculiar galaxy pairs \citep{shamir2014automatic}.}
\label{peculiar_pairs}
\end{figure}

\begin{figure}[h]
\includegraphics[scale=0.95] {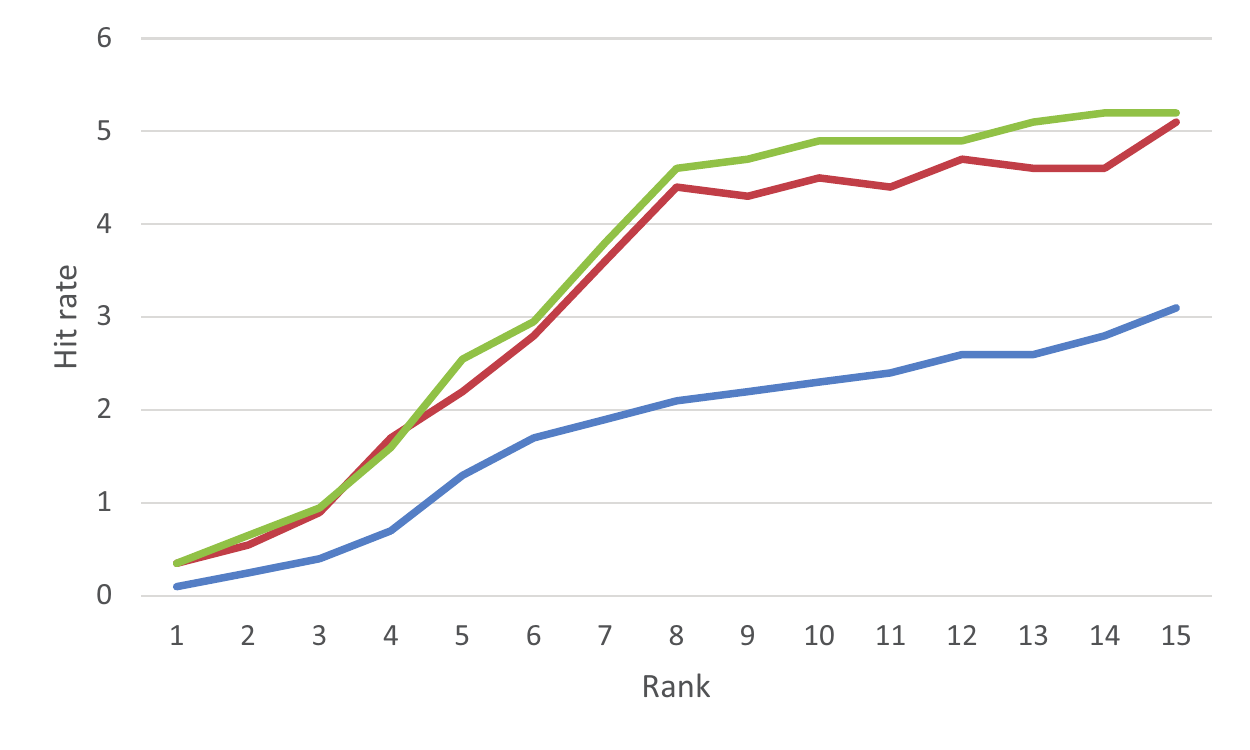}
\caption{Hit rate of the query-by-example algorithm when retrieving 20 peculiar galaxy pairs among 400 galaxies, and the query galaxy is a galaxy pair.}
\label{peculiars_in_400}
\end{figure}

\begin{figure}[h]
\includegraphics[scale=0.75] {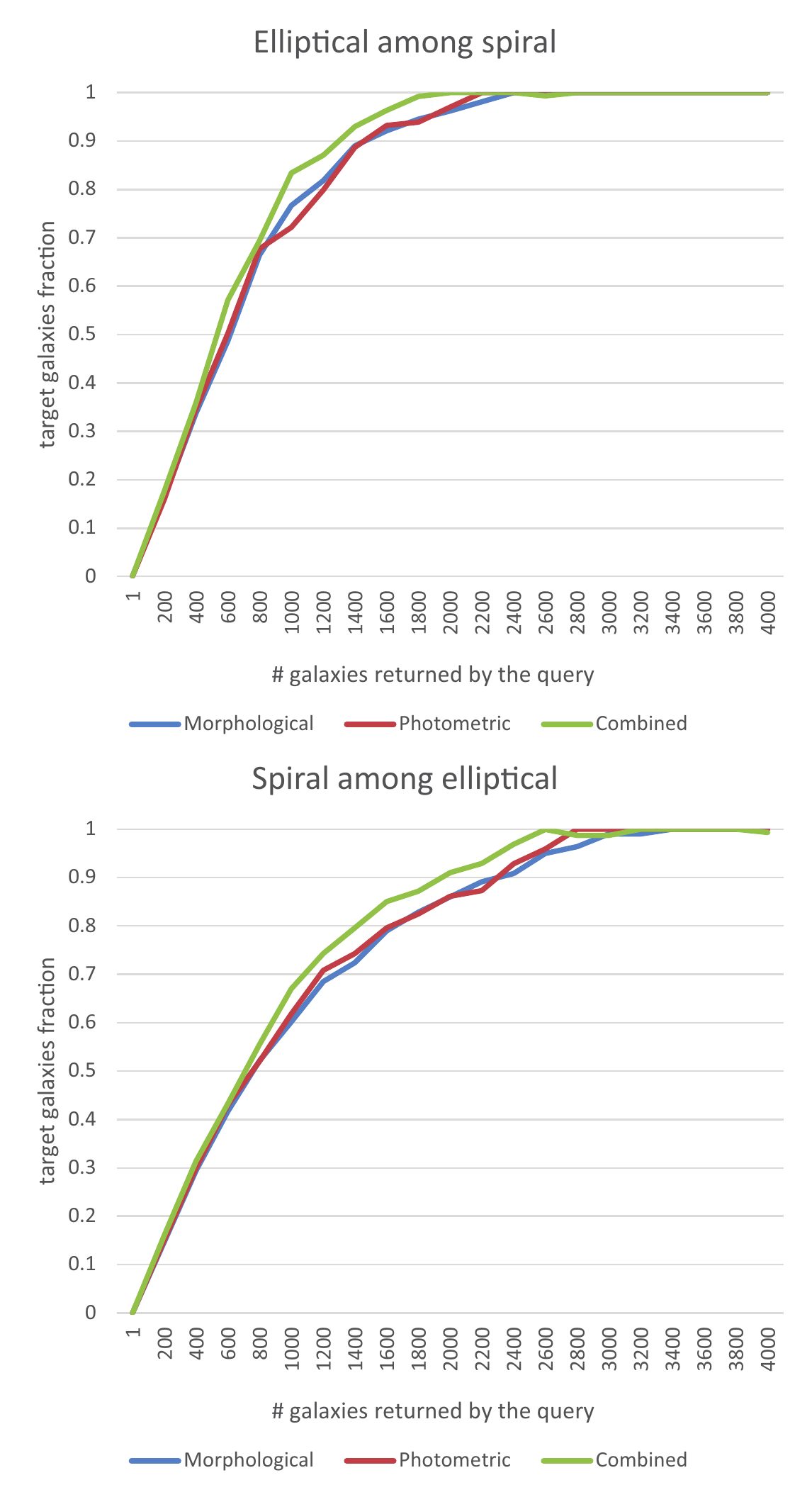}
\caption{The portion of elliptical galaxies (top) and spiral galaxies (bottom) when combining 1,000 target galaxies with 4,000 database galaxies. The Y axis shows the number of the target galaxies (elliptical or spiral) returned by the algorithm divided by the total number of target galaxies.}
\label{completeness}
\end{figure}

The hit rate when using the morphological features, photometric features, and combined feature sets are shown in Figures~\ref{elliptical_in_spiral_100} through~\ref{completeness}. Figure~\ref{elliptical_in_spiral_100} shows the hit rate when using 100 spiral galaxies as the regular galaxies and 10 elliptical galaxies as the peculiar galaxies. As explained in Section~\ref{qbe}, the test is performed multiple times such that in each run different query images are used for testing. The test is performed 100 times such that 10 randomly selected elliptical galaxies are used as the ``peculiar'' galaxies, and one is selected as the query galaxy. The hit rate naturally changes based on the rank, as a larger number of galaxies returned by the query increases the chance that one of them is of the same type as the query galaxy. Similarly, Figure~\ref{spiral_in_elliptical_100} shows the hit rate when attempting to return automatically the $R$ most similar galaxies to a query spiral galaxy among a dataset of 110 galaxies -- 100 elliptical galaxies and 10 spiral galaxies. In both cases the combined feature set that includes both the morphological and photometric features outperforms the performance when using the morphological or photometric features alone.

Another experiment attempted to identify ring galaxies based on a query image. The database galaxies in this experiment were 100 elliptical galaxies, and the query galaxies were 20 ring galaxies used in \citep{shamir2016morphology}. In each run a different ring galaxy is used as the query galaxy, and 10 ring galaxies are combined with the 100 images of elliptical galaxies. Similarly, an additional experiment used the same ring galaxies among 100 spiral galaxies. Figures~\ref{rings_in_ellipticals} and~\ref{rings_in_spiral} show the hit rate of the ring galaxies among the elliptical and spiral galaxies, respectively. In both cases using the combination of morphological and photometric features increased the number of ring galaxies returned by the algorithm given the query ring galaxy image.

To test actual peculiar systems, we combined a set of 20 tidally distorted galaxy pairs with 400 non-peculiar galaxy pairs taken from Sloan Digital Sky Survey. The peculiar galaxy pairs were taken from the catalog of automatically-identified galaxy pairs \citep{shamir2014automatic}, and displayed by Figure~\ref{peculiar_pairs}.

Figure~\ref{peculiars_in_400} shows the average hit rate when using one peculiar galaxy pair as the query galaxy, such that each of the 20 galaxies is used as the query galaxy for each rank. The results show that the photometric features alone provide low hit rate compared to the morphological features and the combined feature sets.

To test the completeness of the query-by-example algorithm, 4,000 elliptical galaxies were combined with 1,000 spiral galaxies, and a spiral galaxy was used as the query image. A similar experiment was done with 4,000 spiral galaxies merged with 1,000 images of elliptical galaxies, and an elliptical galaxy used as the query image in each run. As before, the experiment was repeated such that in each run a different image was used as the query image. The fraction of the target galaxies in the database among the galaxies returned by the query is shown in Figure~\ref{completeness}. As expected, the combined feature set provides more information, allowing the higher frequency of the target galaxies that are similar to the query galaxy.

\begin{figure}[h]
\includegraphics[scale=0.75] {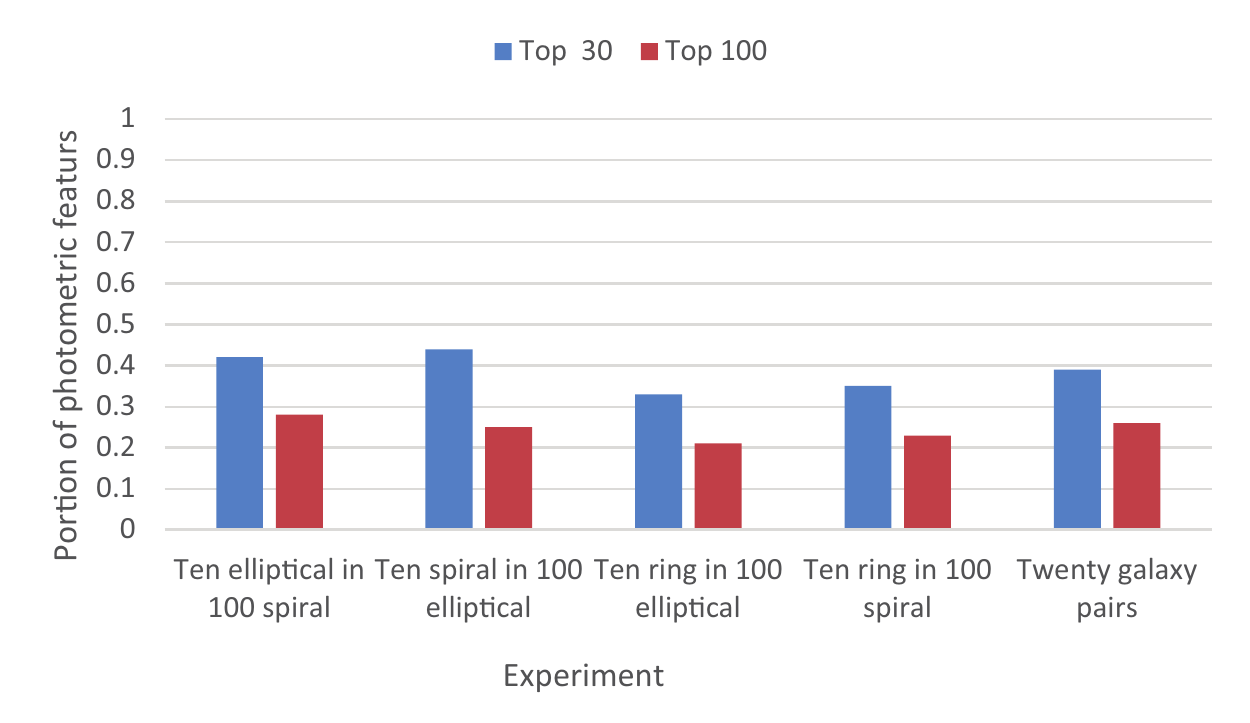}
\caption{Distribution of the photometric and morphological features in each of the experiments.}
\label{feature_distribution}
\end{figure}

Figure~\ref{feature_distribution} shows the distribution of the photometric and morphological features used for each of the experiments. These features are selected by their entropy as a heuristic estimation of their usefulness in identifying similar galaxies to a given query galaxy \citep{shamir2016morphology}. In each experiment different features can be selected based on the data, but the figure shows that in general more morphological features are ranked higher than the photometric measurements. However, the number of morphological features is also much higher than the number of photometric features (2898 compared to 418), so higher noise in the feature selection process is expected to lead to an increased representation of morphological features.

\section{Correlation between photometry and morphology}
\label{correlation}

The set of photometric variables collected by SDSS is clearly not orthogonal, and many of these variables correlate with each other. To test the correlation between the different individual photometric variables and the morphology of the galaxies we performed a regression between the images and each of the photometric variables, and measured the Pearson correlation between the values of each photometric variable and the morphology of the galaxy as measured by the morphological features described briefly in Section~\ref{morph}. That was done by attempting to predict the value of the variable by first training a machine learning system, and then using a test set to compare between the predicted value and the actual value of the variable, such that the value is predicted by using the galaxy image as the input. High correlation between the value predicted when the galaxy image is used as input ans the actual value of the variable indicates a link between the variable and the morphology of the galaxies.

Unlike the classification described in Section~\ref{classification}, the photometric variables are continuous values and not discrete classes, and therefore require different analysis that can correlate the galaxy images with the continuous values. To weigh the numerical image content descriptors by their relevance to a specific photometric attribute, the Pearson correlation between each image feature {\it f} and the continuous variable {\it v} is used as described by Equation~\ref{pearson}

\begin{equation}
W_f= | \frac{1}{N} \sum_{i=1}^{N}   { (\frac{f_i-\overline{f}}{\sigma_f}) \cdot (\frac{v_i-\overline{v}}{\sigma_v}) }  |,
\label{pearson}
\end{equation}
where $W_f$ is the weight assigned to feature {\it f}, and {\it N} is the number of images in the training set. After each image feature is assigned with a weight, the 85\% of the features with the lowest weight are rejected from the analysis. The intuition of this weighting method is that numerical image content descriptors that have higher Pearson correlation with a certain photometry attribute better reflect the morphology that may be associated with it.

The predicted value of a given test image is determined by interpolating the values of its five closest training samples, where the distance between a test image and training images is determined by the weighted Euclidean distance, such that the Pearson correlations are used as weights as described in Equation~\ref{wnn}

\begin{equation}
d=\sqrt{\sum_{f=1}^{|X|} W_f(X_f-Y_f)^2},
\label{wnn}
\end{equation}
where $W_f$ is the assigned Pearson correlation of feature {\it f} computed by Equation~\ref{pearson}, and {\it d} is the computed weighted distance between the test feature vector {\it X} and a training feature vector {\it Y}. The method of correlating an image with a numerical variable is thoroughly described in \citep{shamir2011computer}, and the code is available \citep{shamir2013wnd}.

The experiments were performed such that for each photometric variable 8,000 were selected randomly as a training set, and 2,000 were selected as test set. That was repeated 10 times for each photometric variable such that in each run different galaxies were selected randomly for training and test sets. The Pearson correlations between the value of the actual value of the variable and the predicted value based on the morphology of the galaxy was averaged. A higher correlation suggests that there is a stronger link between the value of the variable and the morphology of the galaxy. Table~\ref{correlations} shows the photometric variables with Pearson correlation higher than 0.1, all of these correlations are statistically significant.

\begin{longtable}[t]{|l||l|}
\hline
Pearson      & Variable   \\
Correlation &                 \\
\hline

0.6592 & u\_r \\
0.6555 & u\_i \\
0.6383 & q\_i \\
0.6373 & q\_z \\
0.6355 & q\_r \\
0.6149 & u\_z \\
0.6117 & u\_g \\
0.5529 & fracDeV\_r \\
0.5348 & fracDeV\_i \\
0.5226 & fracDeV\_g \\
0.5203 & q\_g \\
0.5115 & modelMag\_u \\
0.5101 & fracDeV\_z \\
0.4978 & expPhi\_r \\
0.4964 & expMag\_u \\
0.4926 & deVMag\_u \\
0.489 & deVPhi\_g \\
0.4886 & deVPhi\_r \\
0.4847 & petroMag\_i \\
0.4838 & expPhi\_g \\
0.4817 & expPhi\_i \\
0.4803 & expMag\_i \\
0.4685 & deVRad\_r \\
0.4685 & dered\_u \\
0.4648 & deVPhi\_i \\
0.4625 & deVMag\_r \\
0.4609 & expPhi\_z \\
0.4594 & z \\
0.4546 & deVAB\_i \\
0.4528 & deVPhi\_z \\
0.4528 & dered\_z \\
0.4519 & r \\
0.4505 & modelMag\_g \\
0.4503 & modelMag\_z \\
0.45 & deVMag\_z \\
0.4493 & deVAB\_r \\
0.4483 & deVRad\_g \\
0.4477 & fiberMag\_r \\
0.4471 & deVRad\_i \\
0.4455 & expAB\_r \\
0.445 & expMag\_g \\
0.4436 & u \\
0.4433 & modelMag\_i \\
0.4432 & deVMag\_i \\
0.4402 & petroMag\_r \\
0.4378 & dered\_r \\
0.4362 & expMag\_z \\
0.4359 & deVAB\_g \\
0.435 & i \\
0.4346 & expRad\_z \\
0.4335 & deVMag\_g \\
0.4322 & fiberMag\_z \\
0.4318 & lnLStar\_r \\
0.4284 & fiberMag\_g \\
0.4278 & petroMag\_u \\
0.4265 & expAB\_g \\
0.4233 & expAB\_i \\
0.4218 & petroR90\_g \\
0.4213 & petroMag\_z \\
0.4212 & fiberMag\_i \\
0.4199 & petroMag\_g \\
0.418 & petroR50\_i \\
0.4169 & modelMag\_r \\
0.4134 & deVABErr\_r \\
0.4119 & expMag\_r \\
0.4088 & deVRad\_z \\
0.4083 & lnLDeV\_g \\
0.4081 & fiberMag\_u \\
0.4068 & g \\
0.4023 & dered\_i \\
0.3979 & expRad\_g \\
0.3937 & psfMag\_g \\
0.3927 & modelMagErr\_r \\
0.3926 & dered\_g \\
0.3895 & htmID \\
0.3877 & deVAB\_z \\
0.3856 & expABErr\_r \\
0.3848 & psfMag\_z \\
0.3808 & expAB\_z \\
0.3803 & lnLStar\_i \\
0.3798 & deVMagErr\_r \\
0.3795 & fiberMagErr\_r \\
0.3762 & expRad\_r \\
0.3761 & lnLStar\_g \\
0.376 & expRad\_i \\
0.3721 & psfMag\_u \\
0.3721 & psfMag\_i \\
0.3674 & psfMag\_r \\
0.3628 & fracDeV\_u \\
0.3563 & lnLExp\_z \\
0.3549 & deVMagErr\_i \\
0.3512 & lnLDeV\_r \\
0.3502 & deVPhi\_u \\
0.3495 & lnLStar\_z \\
0.3453 & modelMagErr\_g \\
0.3451 & lnLExp\_i \\
0.3326 & deVABErr\_z \\
0.3317 & texture\_z \\
0.3313 & deVRadErr\_z \\
0.3294 & deVRadErr\_r \\
0.3291 & petroRad\_g \\
0.3231 & lnLExp\_r \\
0.3209 & lnLDeV\_z \\
0.3186 & texture\_i \\
0.3163 & expPhi\_u \\
0.3145 & expMagErr\_z \\
0.31 & deVMagErr\_u \\
0.3096 & petroR90\_r \\
0.3034 & mE1\_r \\
0.3015 & deVRad\_u \\
0.2976 & lnLExp\_g \\
0.2949 & expMagErr\_u \\
0.2887 & petroR50\_g \\
0.2858 & modelMagErr\_z \\
0.2821 & lnLStar\_u \\
0.2806 & modelMagErr\_i \\
0.2799 & petroR50\_r \\
0.2731 & lnLDeV\_i \\
0.2691 & deVRadErr\_i \\
0.2658 & mE1\_g \\
0.2656 & mE2\_r \\
0.2652 & colcErr\_z \\
0.2603 & expRadErr\_z \\
0.2581 & petroRad\_r \\
0.2561 & mRrCc\_r \\
0.2559 & expRadErr\_r \\
0.2531 & deVRadErr\_g \\
0.2498 & mE1PSF\_r \\
0.2474 & deVABErr\_g \\
0.2444 & petroRad\_i \\
0.2334 & u\_u \\
0.233 & mE2\_i \\
0.2297 & deVABErr\_i \\
0.2279 & mE1\_i \\
0.2242 & modelMagErr\_u \\
0.2237 & expAB\_u \\
0.2165 & fiberMagErr\_i \\
0.2067 & deVAB\_u \\
0.2064 & mE1PSF\_g \\
0.2009 & petroR90Err\_u \\
0.1961 & deVMagErr\_z \\
0.1945 & mE2PSF\_r \\
0.1925 & mRrCc\_z \\
0.1924 & nProf\_g \\
0.1912 & expRadErr\_i \\
0.1809 & mRrCcPSF\_r \\
0.1798 & isoPhiGrad\_u \\
0.1794 & expABErr\_z \\
0.1784 & lnLDeV\_u \\
0.1782 & petroR90Err\_i \\
0.1777 & lnLExp\_u \\
0.176 & petroR90\_i \\
0.1743 & mE2PSF\_g \\
0.1742 & mE1PSF\_i \\
0.1742 & mE2PSF\_i \\
0.1712 & mRrCcPSF\_u \\
0.1708 & mRrCcPSF\_g \\
0.1705 & mCr4\_g \\
0.1701 & psfMagErr\_i \\
0.1684 & nProf\_i \\
0.1657 & petroR50\_z \\
0.1638 & mE1PSF\_z \\
0.1635 & texture\_g \\
0.1615 & colcErr\_r \\
0.1609 & isoA\_u \\
0.1608 & mE1PSF\_u \\
0.1566 & isoRowcGrad\_u \\
0.1564 & mE2PSF\_u \\
0.1562 & mRrCcPSF\_z \\
0.1554 & mCr4\_i \\
0.154 & petroRad\_z \\
0.154 & petroRadErr\_u \\
0.1529 & isoColc\_u \\
0.1528 & mRrCcPSF\_i \\
0.1527 & petroR50Err\_r \\
0.1527 & isoRowc\_u \\
0.1524 & isoAGrad\_u \\
0.1515 & uErr\_g \\
0.1489 & petroR90Err\_z \\
0.148 & expMagErr\_i \\
0.1478 & uErr\_i \\
0.1468 & deVMagErr\_g \\
0.1459 & petroR90Err\_g \\
0.1447 & isoAErr\_u \\
0.1398 & isoPhi\_u \\
0.1382 & psfMagErr\_r \\
0.1371 & isoColcGrad\_u \\
0.1361 & deVRadErr\_u \\
0.1345 & texture\_r \\
0.1324 & isoBGrad\_u \\
0.1302 & isoColcErr\_u \\
0.1295 & mCr4PSF\_g \\
0.1282 & petroMagErr\_r \\
0.1248 & mE2\_z \\
0.1241 & mCr4\_r \\
0.124 & isoPhiErr\_u \\
0.1237 & mCr4PSF\_r \\
0.1234 & mRrCcErr\_i \\
0.1225 & uErr\_r \\
0.1212 & qErr\_g \\
0.1203 & psfMagErr\_u \\
0.1178 & mCr4PSF\_z \\
0.1152 & petroR50Err\_i \\
0.1149 & expRadErr\_g \\
0.1134 & q\_u \\
0.1134 & isoB\_u \\
0.1086 & expRad\_u \\
0.1048 & mCr4PSF\_u \\
0.1047 & nProf\_u \\
0.1037 & mCr4\_z \\
0.1029 & mE1\_z \\
0.1018 & mE2E2Err\_g \\
0.1006 & expABErr\_g \\

\hline
\caption{Variables with Pearson correlation between the actual and predicted values greater than 0.1.}
\label{correlations}
\end{longtable}

As the table shows, the variables that had the strongest correlation with the morphology of the galaxies is the Stokes Q and U parameters measured in the different bands. The ``Stokes U' parameter is measured in SDSS by $U=\frac{a-b}{a+b}\sin(2\phi)$, where {\it a} is the major axis, {\it b} is the minor axis of the galaxy, and $\phi$ is the position angle \citep{abazajian2009seventh}. As can be expected, the different magnitude variables also show strong correlation with the morphological descriptors of the galaxies, as well as the position angle variables measured in the different bands. However, the magnitude error also exhibits correlation with the morphology of the images, showing that the error in measuring the magnitude depends on the morphology of the galaxies.

\section{Conclusion}
\label{conclusion}

The increasing importance of autonomous sky surveys and large astronomical databases reinforces the development and application of methods for automatic analysis of astronomical data. Manual analysis of galaxy morphology using crowdsourcing has provided datasets of galaxy morphology that were useful for numerous studies. However, despite the success of these campaigns to recruit a high number of volunteers, these activities did not provide a complete analysis of all galaxies with visible morphology. For instance, the successful Galaxy Zoo 1 campaign \citep{galaxyzoo1} provided a ``superclean'' dataset of less than $7\cdot10^5$ galaxies, which is far smaller than the number of SDSS galaxies with identifiable morphology, and smaller than automatically annotated catalogs of the same digital sky survey \citep{kuminski2016computer}. In the era of LSST, it is clear that full morphological analysis of the galaxies imaged by the future digital sky survey will require automation.

Since many digital sky surveys provide both photometry and image data, these data can be combined to perform a more informative automatic analysis. Here we showed that when using photometry data the performance of the analysis is comparable to analyzing the images directly, and combining morphological and photometric descriptors improves the performance of two pattern recognition tasks -- classification and query-by-example. Nearly all experiments performed in this study showed improved performance when using both morphological and photometric features, although in some of the cases the improvement was marginal. 

While the common photometric measurements computed by digital sky surveys reflect information from the images, these pre-defined standard measurements might not be able to contain all possible information about the morphology of the galaxy. Therefore, additional information provided by applying machine vision methods can improve the automatic analysis tasks. The experiments show that the machine vision analysis adds substantial additional information to tasks such as query-by-example of peculiar galaxies, while making marginal contribution to other tasks such as supervised classification of galaxies annotated  by their morphological types.

As automatic methods are already producing catalogs, the methods described in this paper can be used for tasks such as automatic annotation of galaxies to allow structured queries of the data, as well as query-by-example to identify collections of galaxies that are similar to a query galaxy.

The source code of the method is publicly available \citep{shamir2017udat} through the Astrophysics Source Code Library \citep{allen2012practices}, or at \url{http://vfacstaff.ltu.edu/lshamir/downloads/UDAT}.


\section{Acknowledgments}

This study was supported by NSF grant IIS-1546079.

Funding for the SDSS and SDSS-II has been provided by the Alfred P. Sloan Foundation, the Participating Institutions, 
the National Science Foundation, the US Department of Energy, the National Aeronautics and Space Administration, 
the Japanese Monbukagakusho, the Max Planck Society, and the Higher Education Funding Council for England. The 
SDSS Web Site is http://www.sdss.org/. 
The SDSS is managed by the Astrophysical Research Consortium for the Participating Institutions. The Participating 
Institutions are the American Museum of Natural History, Astrophysical Institute Potsdam, University of Basel, 
University of Cambridge, Case Western Reserve University, University of Chicago, Drexel University, Fermilab, the 
Institute for Advanced Study, the Japan Participation Group, Johns Hopkins University, the Joint Institute for Nuclear 
Astrophysics, the Kavli Institute for Particle Astrophysics and Cosmology, the Korean Scientist Group, the Chinese 
Academy of Sciences (LAMOST), Los Alamos National Laboratory, the Max Planck Institute for Astronomy (MPIA), 
the Max Planck Institute for Astrophysics (MPA), New Mexico State University, Ohio State University, University 
of Pittsburgh, University of Portsmouth, Princeton University, the United States Naval Observatory and the University 
of Washington.

\appendix
\section{Using the code}
\label{using the code}

To meet standard practices of using source code in academic literature \citep{shamir2013practices}, the source code developed and used in this study has been made available to the community through the Astrophysics Source Code Library \citep{allen2012practices,allen2015improving}. The code is also available at \url{http://vfacstaff.ltu.edu/lshamir/downloads/UDAT}. It can be compiled with GNU autotools, and binary files for MS-Windows are also available.

UDAT is a command line tool that can be used with a set of commands, as explained in \citep{shamir2008wndchrm}. In summary, testing the classification accuracy is done by the command: \newline
udat test $<$switches$>$ /path/to/dataset.fit /path/to/report.html \newline

The file ``dataset.fit'' is the file of computed image numerical content descriptors and/or photometric descriptors, and it is also created by UDAT as will be described later in this section. The switches are explained in  \citep{shamir2008wndchrm}, and a brief description is also available when typing ``udat -h''. The file ``report.html'' is an optional file describing the results of the experiment, and it is created automatically by UDAT.
 
To classify a single galaxy the following command can be used: \newline

udat classify $<$switches$>$ /path/to/dataset.fit /path/to/image.tif \newline

Note that the images should be in TIF or PPM format.

The query-by-example can be run in a similar fashion:
udat qbe $<$switches$>$ /path/to/dataset.fit /path/to/query\_image.tif \newline

By default, the command will only return the sample in the dataset that is the most similar to the query sample. To return more than one sample, the switch ``N'' can be used, followed by the number of samples that the query returns. For instance, using the switch ``-N10'' will return the 10 samples in the dataset that are determined by the algorithm to be the most similar to the query sample.

The step of computing the image numerical content descriptors for creating the dataset files (in the example above the file is called ``dataset.fit'') is performed by the following command: \newline

udat compute $<$switches$>$ /path/to/input\_file.cor /path/to/dataset.fit

The command will create the file ``dataset.fit'' from the information specified in the file ``input\_file.cor''. The file ``input\_file.cor'' is a text file of the following format: \newline
/path/to/image1.tif $<$tab$>$ $<$class\_id$>$,photometric\_value1,photometric\_value2,photometric\_value3,... 
/path/to/image2.tif $<$tab$>$ $<$class\_id$>$,photometric\_value1,photometric\_value2,photometric\_value3,... 
/path/to/image3.tif $<$tab$>$ $<$class\_id$>$,photometric\_value1,photometric\_value2,photometric\_value3,... 
   .\newline
   .\newline
   .\newline

UDAT computes the morphological features from each image specified by the full path to an image file name, and then adds the photometric features values to the feature vector of each image to create the combined dataset of morphological and photometric features.

It should be noted that the process of computing the numerical content descriptors is computationally intensive \citep{shamir2008wndchrm} that normally requires a computing cluster \citep{shamir2014automatic,kuminski2016computer}. The resulting output file can also be large, and each sample can add $\sim$30 kilobyte to the size of the file.




\bibliographystyle{elsarticle-harv}
 \biboptions{authoryear}
\bibliography{ms}




\end{document}